\documentclass[conference]{IEEEtran}
\usepackage{graphicx}
\usepackage{siunitx}
\usepackage{pdfpages}       
\usepackage{amsmath}
\usepackage{amssymb}   
         
\makeatletter
\let\MYcaption\@makecaption
\long\def\@makecaption#1#2{%
	\vskip\abovecaptionskip
	\sbox\@tempboxa{#1: #2}%
	\ifdim \wd\@tempboxa >\hsize
	#1: #2\par
	\else
	\global \@minipagefalse
	\hb@xt@\hsize{\hfil\box\@tempboxa\hfil}%
	\fi
	\vskip\belowcaptionskip}
\makeatother

\usepackage[font=footnotesize]{subcaption}

\makeatletter
\let\@makecaption\MYcaption
\makeatother

\begin{document}
\title{An Experimental Study on \\Airborne Landmine Detection Using a \\Circular Synthetic Aperture Radar}
\author{\IEEEauthorblockN{
		Markus Schartel\IEEEauthorrefmark{1}$^,$\IEEEauthorrefmark{4},
		Ralf Burr\IEEEauthorrefmark{2},
		Rik B\"ahnemann\IEEEauthorrefmark{3},
		Winfried Mayer\IEEEauthorrefmark{4}, and
		Christian Waldschmidt\IEEEauthorrefmark{1}}
	\IEEEauthorblockA{\IEEEauthorrefmark{1}Institute of Microwave Engineering, Ulm University, 89081 Ulm, Germany}
	\IEEEauthorblockA{\IEEEauthorrefmark{2}Ulm University of Applied Sciences, 89075 Ulm, Germany}
	\IEEEauthorblockA{\IEEEauthorrefmark{3}Autonomous Systems Lab, ETH Z\"urich, 8092 Z\"urich, Switzerland}
	\IEEEauthorblockA{\IEEEauthorrefmark{4}Endress+Hauser SE+Co. KG, 79689 Maulburg, Germany}
	Email: markus.schartel@alumni.uni-ulm.de}
\maketitle
\begin{abstract}
	Many countries in the world are contaminated with landmines. Several thousand casualties occur every year. Although there are certain types of mines that can be detected from a safe stand-off position with tools, humanitarian demining is still mostly done by hand. As a new approach, an unmanned aerial system (UAS) equipped with a ground penetrating synthetic aperture radar (GPSAR) was developed, which is used to detect landmines, cluster munition, grenades, and improvised explosive devices (IEDs). The measurement system consists of a multicopter, a total station, an inertial measurement unit (IMU), and a frequency-modulated continuous-wave (FMCW) radar operating from \SIrange{1}{4}{\GHz}. The highly accurate localization of the measurement system and the full flexibility of the UAS are used to generate 3D-repeat-pass circular SAR images of buried anti-personnel landmines. In order to demonstrate the functionality of the system, 15 different dummy landmines were buried in a sandbox. The measurement results show the high potential of circular SAR for the detection of minimum metal mines. 11 out of 15 different test objects could be detected unambiguously with \si{cm}-level accuracy by examining depth profiles showing the amplitude of the targets response over the processing depth.
\end{abstract}
\vspace*{3mm}
\begin{IEEEkeywords}
	Anti-personnel mine; frequency-modulated continuous-wave radar (FMCW); ground penetrating radar (GPR), synthetic aperture radar (SAR), unmanned aerial system (UAS).
\end{IEEEkeywords}
\IEEEpeerreviewmaketitle
\section{Introduction}
The rapid progress in the development of unmanned aerial systems (UAS) opens up a wide range of new applications~\cite{Huegler2018}. A major advantage of UAS's is the capability to explore inaccessible or dangerous areas from a safe distance. Humanitarian demining, for example, exposes people to serious risks in their daily work. Equipped with a demining probe, a hand-held metal detector (MD), and/or a ground penetrating radar (GPR), deminers work in close proximity to the threat~\cite{daniels2004ground}. In recent years, several different approaches and sensor principles for the UAS-based detection of mines have been investigated. The optical detection of partially buried mines and cluster munitions is investigated in \cite{Castiblanco2014,droneandbutt2018}. Cameras are well suited for the detection of surface-laid threats, but in the case of buried or hidden mines, these types of sensors fail. The authors of \cite{Gavazzi2016} use fluxgate vector magnetometers mounted on a UAS to detect unexploded ordnances (UXO). However, magnetometers as well as metal detectors cannot detect modern minimum metal mines in a reliable way. A GPR is a measuring device, which is able to generate subsurface images of minimum metal mines. In recent years, much research has therefore been done in the field of UAS-based detection of mines using a down-looking radar \cite{Colorado2017,Fasano2017,Fernandez2018,Garcia-Fernandez2019}. However, a down-looking radar has the disadvantage that the potential minefield has to be scanned line by line, resulting in a low area throughput. In addition, due to the strong ground reflection it is very difficult to detect objects buried just below the surface. To overcome these issues, a side-looking radar can be employed. A long synthetic aperture enables the generation of high-resolution 2D-radar images. However, using a linear aperture, there is an ambiguity problem in range direction and depth. This can be solved by either a nonlinear motion trajectory of the UAS, e.g. a circular synthetic aperture radar (CSAR), or by cross-track interferometric SAR (InSAR), or repeat pass tomography~\cite{Reigber2000,Krieger2010}. Due to the flexibility and highly accurate absolute localization of the used measurement system, both procedures and the combination of them can be performed. The system, the radar, and the antennas have been published in \cite{Schartel2017}, \cite{Burr2018a}, and \cite{Burr2018b}, respectively. The detection of tripwires using this system is investigated in \cite{Schartel2019a}. First subsurface linear SAR (LSAR) measurements of realistic landmine dummys are shown in~\cite{Heinzel2019}.

The UAS-based ground penetrating synthetic aperture radar (GPSAR) system is now able to record radar data from arbitrary trajectories coherently, and for the first time this unique feature is used to detect buried landmines. This paper examines the coherent superposition of data from circular apertures at different heights. Depth profiles are used to show that deeply buried anti-tank mines as well as anti-personnel mines located just below the surface can be detected using a side-looking radar.

This paper is organized as follows. Section~II briefly describes the signal processing chain and the theoretically achievable 3D-resolution. The measurement system and the motion compensation are presented in Section~III. Section~IV presents GPSAR measurements of buried landmines using a circular aperture. Conclusions are drawn in Section~V.
\section{Signal Processing and Resolution}
For image generation a back-projection algorithm is used. In simplified form, the complex value of the pixel $P$ at the location $(x_0,y_0,z_0)$ results from \cite{Zaugg2015}
\begin{align}
P(x_0,y_0,z_0) = \sum_{n=1}^{N}S_n\exp(-\mathrm{j}\phi_{x_0,y_0,z_0})\,,
\end{align}
where $N$ is the number of chirps, $S$ the interpolated and range-compressed radar signal, and $\phi$ the expected phase for the pixel. In this paper the FMCW radar data are range-compressed and interpolated by performing a Hann-windowed fast Fourier transform with 16-fold zero-padding. The phase $\phi_{x_0,y_0,z_0}\,$=$\,2\pi f_0 \Delta t\,$$-$$\pi K\Delta t$ depends on the start frequency $f_0$, the slope of the chirp $K$, and the round-trip time $\Delta t\,$=$\,2(r_0/c_0+r_1/c_1)$. $r_0$ and $r_1$ are the propagation path lengths in the air and the soil from the phase center of the antenna $(x_a,y_a,z_a)$ to the pixel $(x_0,y_0,z_0)$ and can be calculated using the law of refraction \cite{Heinzel2016}. The velocity of propagation $c_1\,$=$c_0/\sqrt{\epsilon_r}$ depends on the speed of light $c_0$ in free space and the  relative permittivity $\epsilon_r$ of the soil. $\epsilon_r$ is assumed to be constant in the area of interest. A 3D-SAR image can be generated by processing the data for different focus planes~$z_0$. In the following, $z_0\,$=$\,$0 is defined as the plane in which the transition from air to soil is located. If $z_0\,$$\geq\,$0, the image is focused on a plane above the reference plane and if $z_0\,$$<\,$0, subsurface processing is performed. 

In CSAR the resolution in $z$-direction depends on the angular persistence of the target \cite{Moore2007}. For targets having an angular persistence greater than \SI{22.5}{\degree} in the $xy$-plane, the resolution in the $z$-direction~$\delta_z$ is close to the resolution defined by the radar bandwidth $\delta_r\,$=$\,\frac{c_0}{B}$, i.e., \cite{Ponce2014}
\begin{align}
\delta_z = \frac{4}{\sqrt{2\pi}\cos{\theta_i}}\frac{c_0}{2B}
\end{align}
and depends on the incidence angle $\theta_i$. Maximum resolution in cross-range $\delta_{cr}$ and ground-range $\delta_{gr}$ is achieved for isotropic targets with an angular persistence of \SI{360}{\degree} and is \cite{Ponce2014} 
\begin{align}
\delta_{cr} = \delta_{gr} = \frac{c_0}{4(f_0+B)\sin{\theta_i}}\,.
\end{align}

For the radar parameters and the geometry presented in Section IV the resolution in the $xy$-plane is estimated to be in the range of \SIrange{2}{4}{\cm} for a target visible with an angular persistence of \SI{360}{\degree} in the area of interest. The resolution in $z$-direction is estimated to be in the range of \SIrange{8}{10}{\cm} in air and \SIrange{3}{4}{\cm} in soil.
\section{Measurement System and Motion Compensation}
\begin{figure}[tbp]
	\centering
	\includegraphics[width=\linewidth]{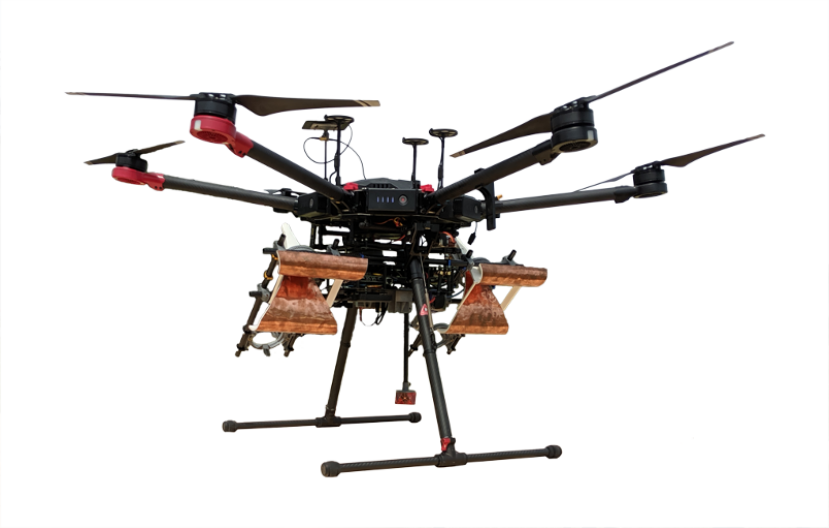}
	\caption{UAS equipped with the \SI{1}{\GHz}$-$\SI{4}{\GHz} FMCW radar and two 3D-printed horn antennas.}	
	\label{fig:measurement_system}
\end{figure}

In Fig.~\ref{fig:measurement_system} a picture of the UAS is shown. The main components are a bistatic frequency-modulated continuous-wave (FMCW) radar, two lightweight horn antennas, an inertial measurement unit (IMU), and an optical prism. As explained in detail below, the position of the prism is tracked by a total station (tachymeter) and linked to the IMU to calculate the position of the antennas. The data are stored, synchronized using the pulse-per-second (PPS) signal of the GNSS, and processed offline. The system parameters are given in Table~\ref{tab:system}.

\begin{table}[b]
	\caption{System parameters of the FMCW GPSAR, the antennas, the total station, and the IMU.}\label{tab:system}
	\begin{tabular}{ll}
		\textbf{Radar:}\\
		Frequency range: & \SI{1}{\GHz}$-$\SI{4}{\GHz}\\
		Transmit power: & $15\,$dBm                   \\
		Chirp duration: & \SI{1024}{\micro\s}		  \\
		Sampling frequency: & \SI{4}{\MHz}		      \\
		Chirp repetition frequency: & \SI{30}{\Hz}    \\
		Sampling distance ($v$=\SI{0.4}{\m/\second}): \SI{13.3}{\mm}\vspace{1.25mm}\\ 
		\textbf{Antennas:}\\         
		Gain:                              & 6$\,$dBi\\
		Vertical \SI{3}{\dB} beamwidth:    & \SI{60}{\degree}\\
		Horizontal \SI{3}{\dB} beamwidth:  & \SI{50}{\degree}\vspace{1.25mm}\\
		\textbf{Tachymeter:}\\
		Distance accuracy:                & \SI{4}{\mm}\\
		Angular accuracy: & \SI{0.00056}{\degree} (\ang{;;2})\\
		Measurement rate:                 & \SI{20}{\Hz}\vspace{1.25mm}\\	
		\textbf{IMU:}\\
		Angular random walk:   & 0.66\si{\degree}/$\sqrt{\text{h}}$\\
		Velocity random walk: & 0.11$\,$m/s/$\sqrt{\text{h}}$\\
		Measurement rate:     & \SI{200}{\Hz}    \\	
	\end{tabular}
\end{table}

A critical component when calculating the round-trip time $\Delta t$ is a precise estimate of the antenna phase center positions along the synthetic radar aperture in space~$(x,y,z)$. 
\begin{figure*}[t]
	\centering
	\includegraphics{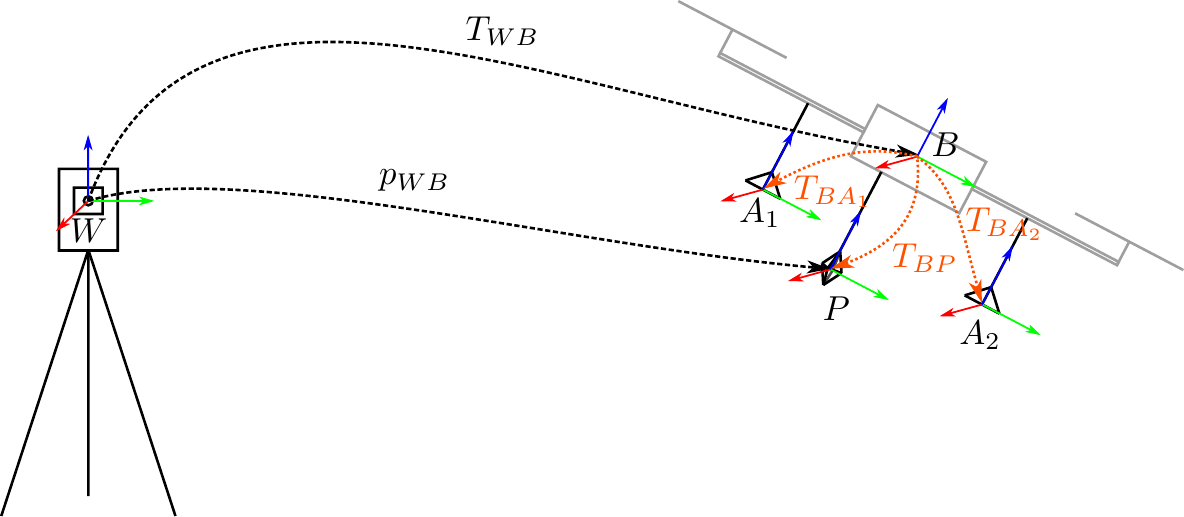}
	\caption{Determination of the phase center positions of the radar antennas $A_i$ by fusion of high-precision total station measurements and high-rate IMU data.}
	\label{fig:estimation_model}
\end{figure*}   

Figure~\ref{fig:estimation_model} sketches the state estimation geometry. The total station tracks the position $p_{WP}$ of the prism $P$ attached to the UAS in the world coordinate frame $W$.
Additionally, the UAS body rates $\omega_{B}$ and linear accelerations $a_B$ in the body frame $B$ are measured using an IMU. These two sensor modalities have to be fused to obtain a six-degree-of-freedom estimate $T_{WB}$ (postition and orientation) of the UAS body.
\begin{figure*}[t]
	\centering
	\includegraphics{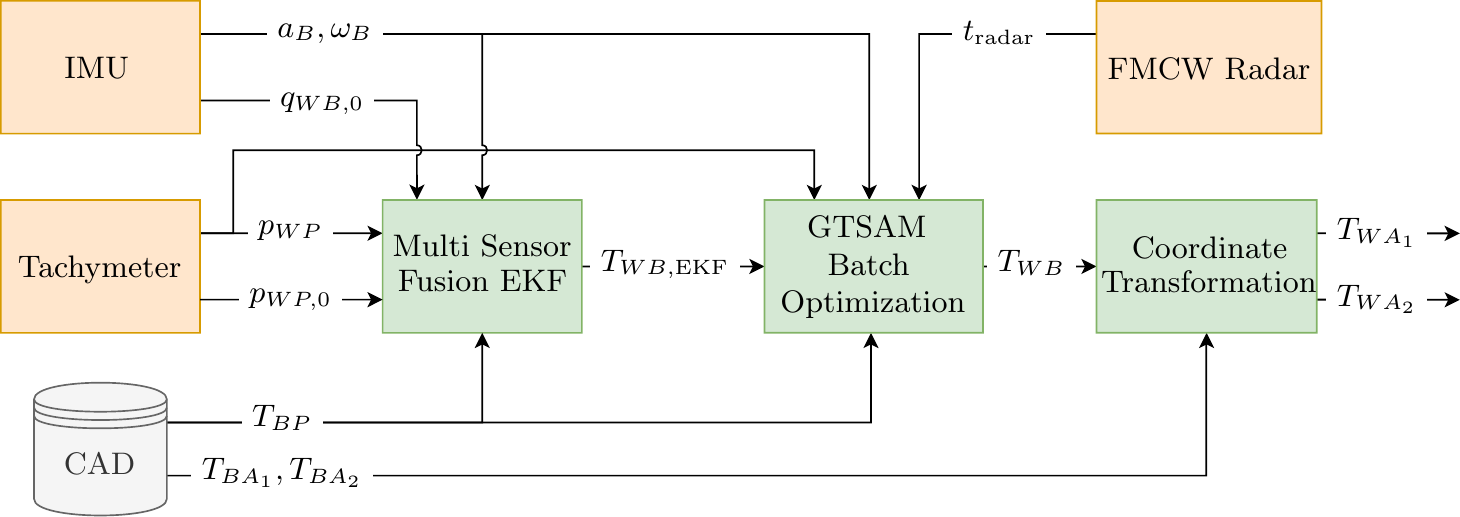}
	\caption{A three-step state estimation pipeline to fuse total station and IMU data.}
	\label{fig:estimation_pipeline}
\end{figure*}

Figure~\ref{fig:estimation_pipeline} shows the implemented three-step state estimation pipeline in detail. In the first step, a solution of the UAS position and orientation $T_{WB,\text{EKF}}$ is calculated using an extended Kalman filter (EKF) \cite{ref-msf}.  Initially, the UAS orientation $q_{WB,0}$ inferred from the IMU and the magnetometer at take-off and the first total station measurement $p_{WB,0}$ initialize the filter. After initialization, the filter integrates the IMU measurements to obtain a high-rate estimate of the UAS position and orientation. This estimate is corrected for drift by the absolute position measurements of the total station. In order to fuse the prism position measurements they are transformed into body frame coordinates using the static homogeneous transform $T_{BP}$, which can be obtained from the Computer-Aided-Design (CAD) model of the measurement system. In a second step, a Georgia Tech Smoothing and Mapping (GTSAM) batch optimization creates an improved and smooth estimate as it takes into account the full measurement history~\cite{ref-gtsam}. As input the initial EKF solution, the IMU and total station measurements, the prism calibration, and the radar time stamps $t_\mathrm{radar}$ are used. Additionally, the batch optimization integrates the IMU to match the radar measurement time stamps. Finally, the antenna positions are calculated based on the offset between body frame and antenna phase centers. The phase center positions and orientations $T_{WA_i}$ can be calculated as
\begin{align}
T_{WA_i} = T_{WB} T_{BA_i}\,,
\end{align}
where $T_{BA_i}$ is the static transformation from the base frame to the antenna frame $i=\{1,2\}$ obtained from the CAD model.
\section{Measurement Setup and Results}
\begin{figure*}[t]
	\centering
	\begin{subfigure}[t]{\linewidth}
		\centering
		\includegraphics[width=\linewidth]{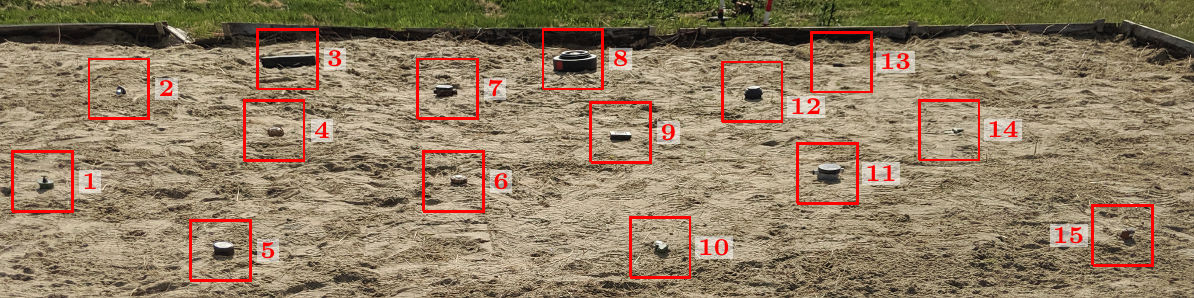}
		\caption{A photo taken before the targets were buried.}
		\label{fig:photo_setup}
	\end{subfigure}
	\\
	\vspace*{2mm}
	\begin{subfigure}[t]{\linewidth}
		\centering
		\includegraphics[width=\linewidth]{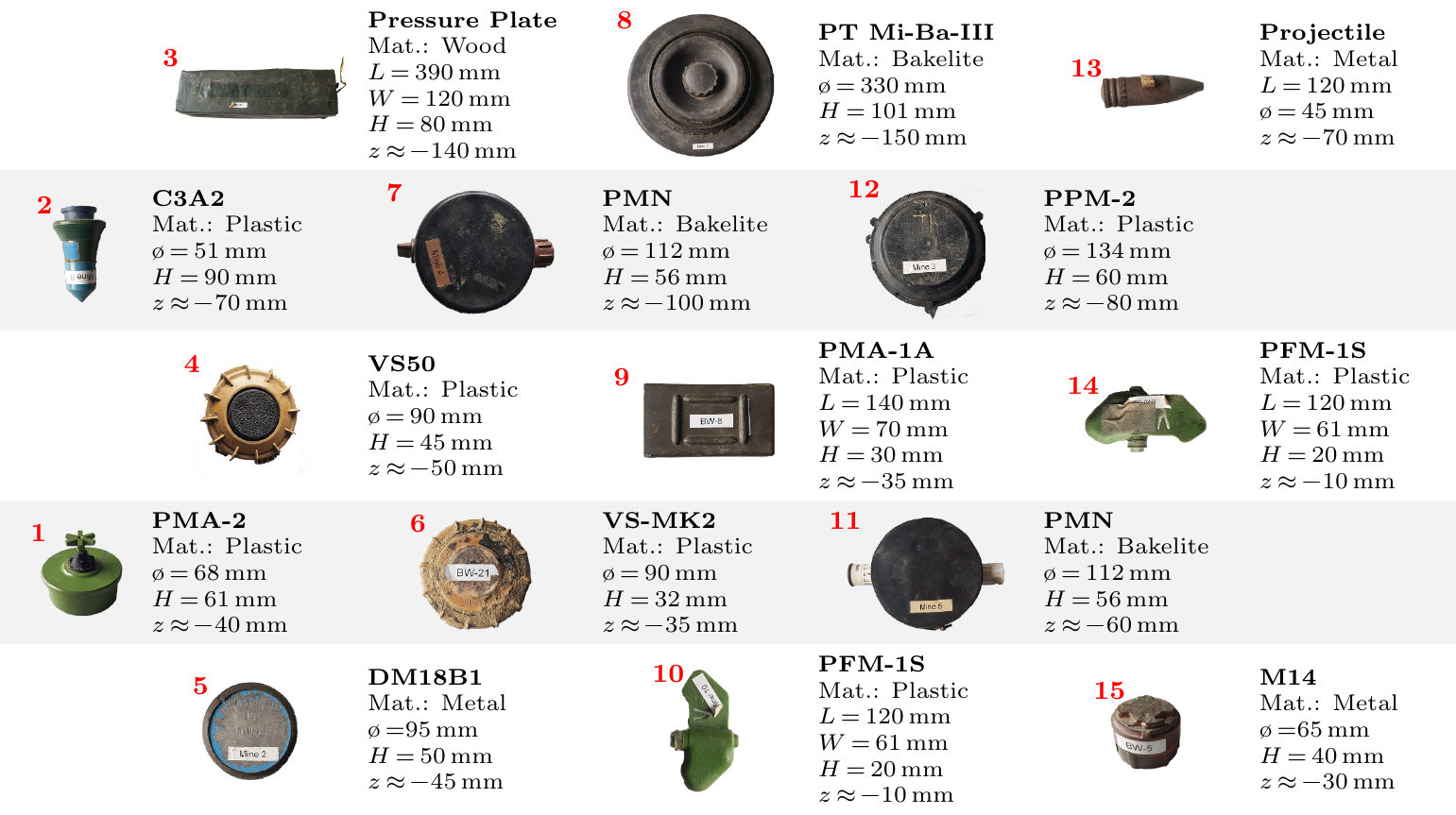}
		\caption{Targets.}	
		\label{fig:measurement_setup}
	\end{subfigure}
	\caption{(\subref{fig:photo_setup}) A photo of the setup consisting of 12 different anti-personnel landmines, one anti-tank mine, a projectile, and a pressure plate and (\subref{fig:measurement_setup}) an overview of the used targets, including their dimensions~\cite{king2018jane} (material Mat., length $L$, width $W$, height $H$, and diameter \o) and the depth $z$ measured from the center of the mine to the surrounding sand surface.}
\end{figure*}

A photo taken before the targets were buried is shown in Fig.~\ref{fig:photo_setup}. The test targets, their dimensions, and an approximate burial depth $z$ are shown in Fig.~\ref{fig:measurement_setup}. The depth refers to the distance from the center of the mine to the surrounding sand surface and is a rough reference value, in which focus plane the target is located. The moisture of the sand was between 12\% and 18\%. A total of 15 different test objects were selected, including 13 anti-personnel landmines. The remaining objects are a wooden pressure plate (target~3), a plastic anti-tank mine (target~8), and a metal projectile (target~13). The mines contain a special silicone as an explosive substitute. The targets were buried as realistically as possible, e.g., targets~1 (PMA-2) and 2 (C3A2) were buried in such a way that the detonators were directly at surface level, targets~10 and 11 (PFM-1S) were only slightly covered with sand, while the anti-tank mine (PT Mi-Ba-III) was buried at a depth of~\SI{-150}{\mm}.

The CSAR measurements were performed around the center of the minefield. In total six apertures in an approximate height from \SIrange{2.5}{5}{\m} with a radius of \SI{7.75}{\m} are evaluated. The trajectories were flown automatically. The control is based on GPS coordinates ($x$,$y$), and a radar altimeter is used to control the altitude above ground $z$. The speed of the UAS was set to \SI{0.4}{\m/\second}. Since no digital elevation model with \si{cm}-level accuracy is available, the detailed topography of the test field is neglected, and a flat surface in the $xy$-plane is used as reference plane. This leads to an error between the locally measured depth of each target in relation to the reference plane in the \si{cm}-range. Furthermore, the assumed position of the interface between air and ground does not necessarily correspond to the real topography.

The data is processed from $z\,$=$\,$\SI{100}{\mm} above to $z\,$=$\,$\SI{-200}{\mm} below the reference plane with a step width of $\Delta z\,$=$\,$\SI{5}{\mm}. If the sign of $z$ is negative, subsurface processing is performed. As mentioned above, the permittivity is assumed to be constant and set to $\epsilon_r\,$=$\,8$ based on an empirical value. Due to the high-precision localization of the UAV, no autofocus algorithm is applied. 

\begin{figure*}[t]
	\centering
	\begin{subfigure}[t]{\linewidth}
		\centering
		\includegraphics[width=0.8\linewidth,height=10cm]{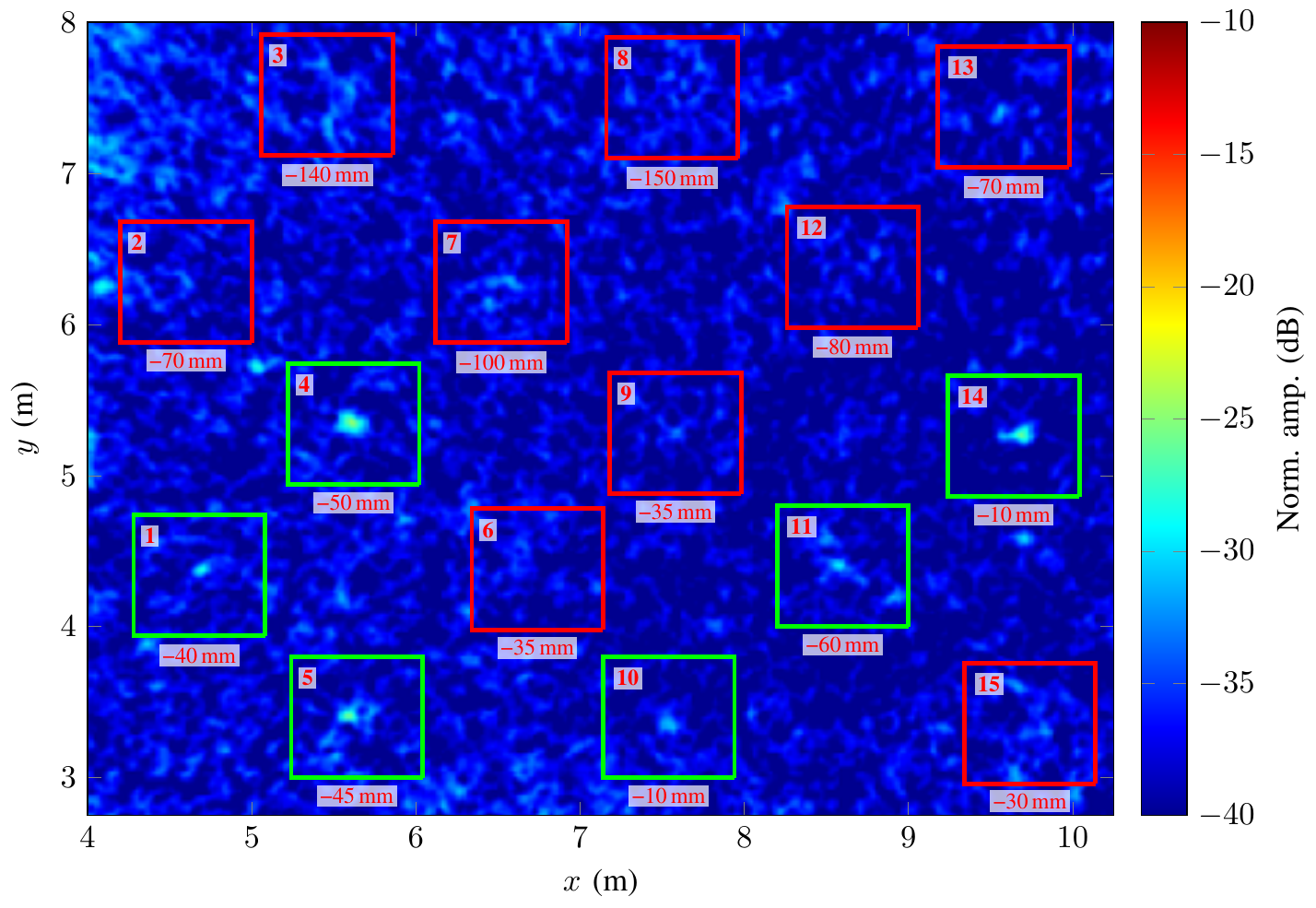}
		\caption{$z\,$=$\,$\SI{-30}{\mm}.}
		\label{fig:csar_03}
	\end{subfigure}
	\\
	\vspace*{2mm}
	\begin{subfigure}[t]{0.47\linewidth}
		\centering
		\includegraphics[width=\linewidth,height=6.5cm]{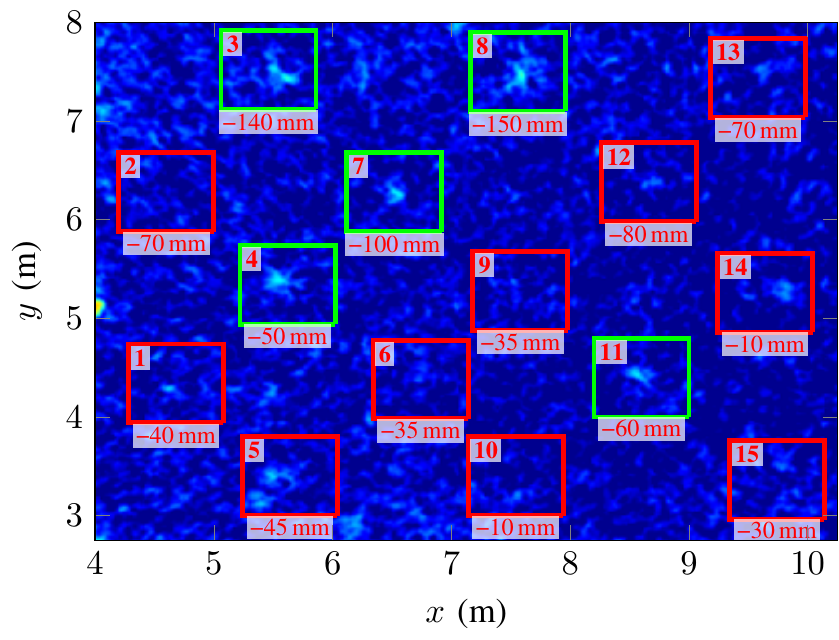}
		\caption{$z\,$=$\,$\SI{-70}{\mm}.}\label{fig:csar_07}
	\end{subfigure}%
	\hfill
	\begin{subfigure}[t]{0.47\linewidth}
		\centering
		\includegraphics[width=\linewidth,height=6.5cm]{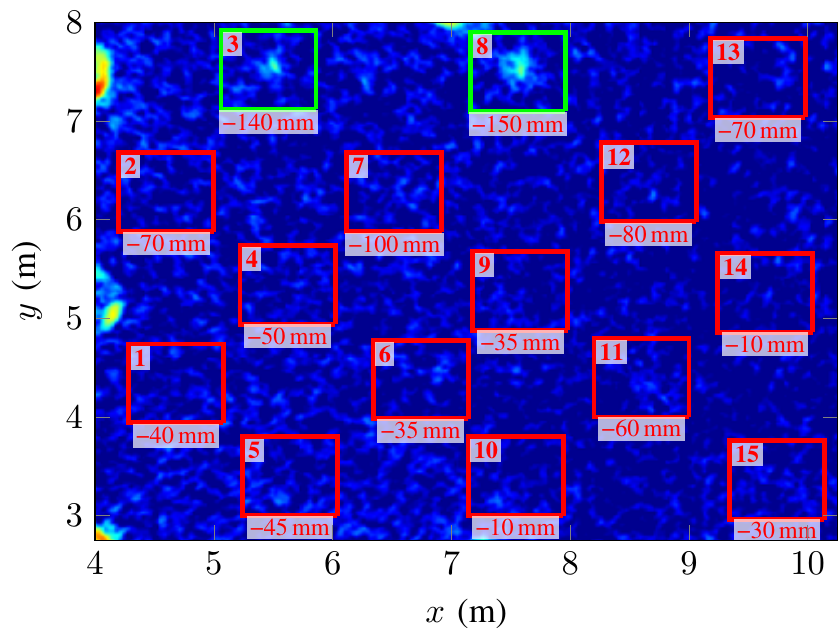}	
		\caption{$z\,$=$\,$\SI{-130}{\mm}.}\label{fig:csar_13}
	\end{subfigure}%
	\caption{GPSAR images for the three focus planes (\subref{fig:csar_03}) $z\,$=$\,$\SI{-30}{\mm},  (\subref{fig:csar_07}) $z\,$=$\,$\SI{-70}{\mm}, and (\subref{fig:csar_13}) $z\,$=$\,$\SI{-130}{\mm} (horizontal polarisation). The green and red boxes indicate the positions of the buried targets. The measured depth $z$ of the targets is given below these boxes.}\label{fig:csar_hh_03_13}
\end{figure*}

Figure~\ref{fig:csar_hh_03_13} shows the result of the coherent superposition of the six SAR images for the measurements with horizontal polarization and three different focus planes. 
The positions of the target reflections are marked with green and red boxes to highlight them. In the focus plane $z\,$=$\,$\SI{-30}{\mm} only the reflection of the targets located on or just below the surface can be seen, e.g., targets 1, 4, 5, 10, 11, and 14 (green boxes in Fig.~\ref{fig:csar_03}). Contrary to expectations, the targets 6, 9, and 15 are not visible in this focus plane using the horizontal polarisation. By changing the focus plane to $z\,$=$\,$\SI{-70}{\mm} or $z\,$=$\,$\SI{-130}{\mm}, targets located on or just below the surface fade out, and the deeper buried targets~3 and 8 become visible (green boxes in Fig.~\ref{fig:csar_07} and~\subref{fig:csar_13}). In addition to targets~6, 9 and 15, target~2 is not visible in these images. Furthermore, Fig.~\ref{fig:csar_13} shows the defocused circular reflection of surface placed reflectors distributed around the test field ($x\,$=$\,$\SI{4}{\m} and $y\,$=$\,\{\SI{2.75}{\m},\SI{5.2}{\m},\SI{7.5}{\m}\}$). 

\begin{figure}[t]
	\centering
	\includegraphics[width=\linewidth,height = 6.25cm]{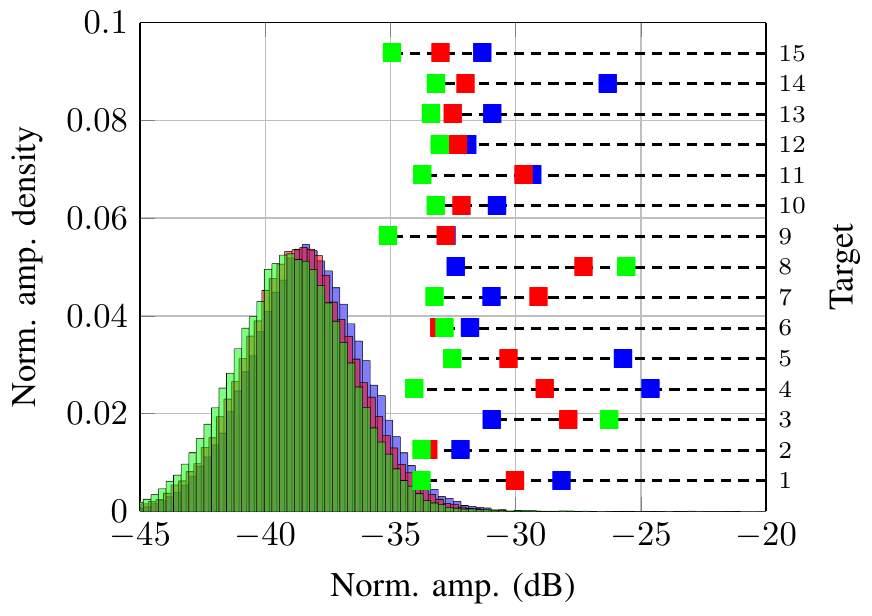}
	\caption{Histograms of the data shown in Fig.~\ref{fig:csar_hh_03_13} and the extracted amplitude of the target response for the focus plane $z\,$=$\,$\SI{-30}{\mm} (\textcolor{blue}{$\blacksquare$}), $z\,$=$\,$\SI{-70}{\mm}~(\textcolor{red}{$\blacksquare$}), and $z\,$=$\,$\SI{-130}{\mm} (\textcolor{green}{$\blacksquare$}).}\label{fig:hist}
\end{figure}

In Fig.~\ref{fig:hist} the histograms depicting the amplitude densities of the GPSAR images are shown. Furthermore, the amplitude for each target is extracted using a maximum search and listed in the histogram. For this purpose, the local maximum within the boxes shown in Fig.~\ref{fig:csar_hh_03_13} is used, regardless of whether the target is visible or not.

\begin{figure}
	\centering
	\includegraphics[width=0.5\linewidth,height=6.25cm]{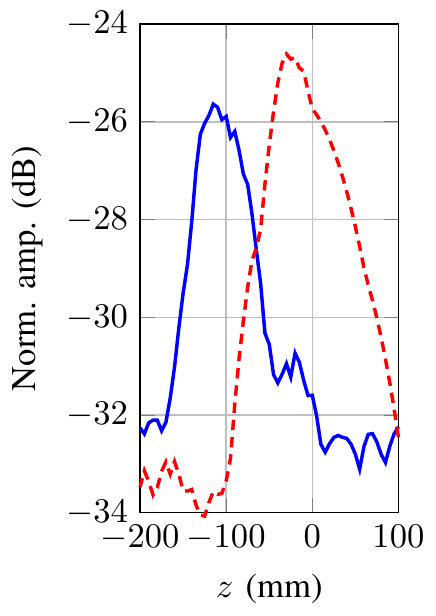}
	\caption{Amplitude versus depth of target~3 (pressure plate, $z\,$$\approx$$\,$\SI{-140}{\mm}, solid blue line) and of target~4 (VS50, 	$z\,$$\approx$$\,$\SI{-50}{\mm}, dashed red line).}
	\label{fig:ampdepth34}
\end{figure}

Both the histograms and the extracted amplitudes depend on the focus plane. Due to the low number of targets and the strong defocusing of surface clutter, the mean value of the Gaussian distribution of the amplitude density decreases at deeper focus planes. In the case of $z\,$=$\,$\SI{-30}{\mm} (\textcolor{blue}{$\blacksquare$}) the amplitudes for the near-surface targets (e.g. targets~4, 5, and 14) are close to their maximum value. In contrast, the amplitude of deeper buried objects (e.g. targets 3 and 8) is greater at a deeper focus plane $z\,$=$\,$\SI{-130}{\mm} (\textcolor{green}{$\blacksquare$}). Target~7 has its maximum amplitude somewhere between these two focal planes (\textcolor{red}{$\blacksquare$}). In Fig.~\ref{fig:ampdepth34} the extracted amplitude of the target response versus the depth $z$ is shown, using target~3 (pressure plate, $z\,$=$\,$\SI{-140}{\mm}) and target~4 (VS50, $z\,$=$\,$\SI{-50}{\mm}) as examples.
For both targets, a variation of the amplitude of more than \SI{6}{\dB} is achieved by varying the focus plane $z$. The unsymmetrical shape of the curves can be explained by the fact that the mines are extended targets with several scattering centers. Although the permittivity of the soil was taken as constant based on an empirical value and the true topography was unknown and neglected, the depth of the targets could be determined with centimeter precision.  

\begin{figure}[h]
	\centering
	\includegraphics{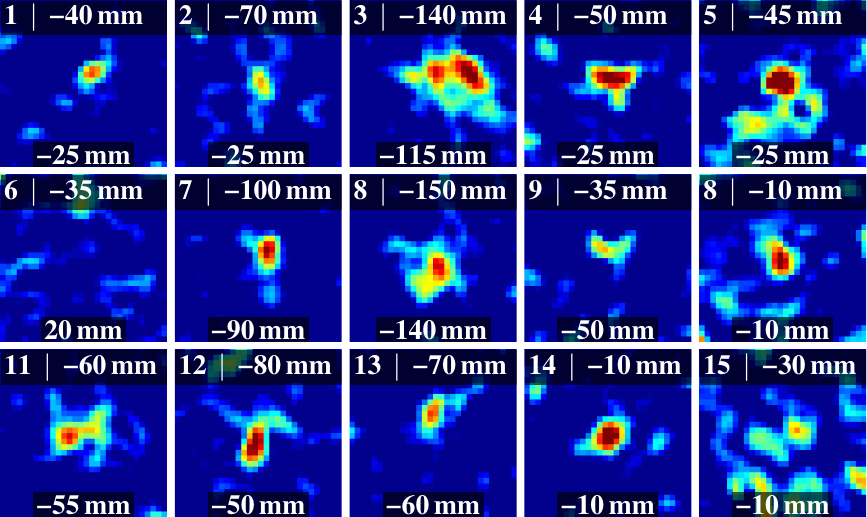}
	\\
	\includegraphics{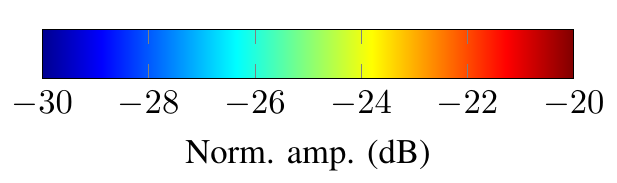}
	\caption{GPSAR image sections of all targets at the focus plane $z$ with the highest amplitude (vertical polarisation). The target number and the measured depth (see Fig.~\ref{fig:measurement_setup}) are at the top, and the focus plane is indicated at the bottom of each image. The dimensions of the image sections are \SI{0.5}{\m}$\,\times\,$\SI{0.5}{\m}.}
	\label{fig:csar_vv}
\end{figure}

\begin{figure}
	\centering
	\includegraphics[width=\linewidth]{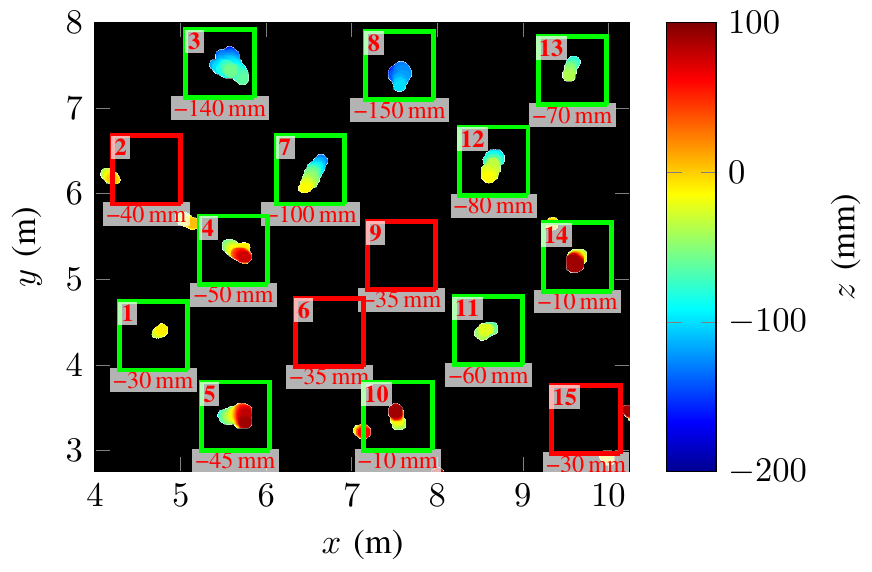}
	\caption{Detected targets using a 2D-CA-CFAR algorithm (vertical polarisation). The depth of the targets $z$ is given below the boxes.}\label{fig:vv_det}
\end{figure}

Figure~\ref{fig:csar_vv} shows the target responses for measurements with vertical polarisation.
The dimensions of the image sections are \SI{0.5}{\m}$\,\times\,$\SI{0.5}{\m} and correspond to the boxes shown in Fig.~\ref{fig:csar_hh_03_13}. The GPSAR image sections are displayed for the depth $z$ at which the highest amplitude is reached. With the exception of target~6, a local maximum is visible in the middle of each of these image sections. In addition, a lot of clutter is visible in the area of target~15, which makes later detection more difficult. As mentioned above, the deviation between the measured depth and the focus plane with maximum amplitude is due to the neglect of the true topography and the fact that the mines are extended objects with multiple scattering centers. For target detection a simple 2D-cell-averaging (CA) constant-false-alarm-rate (CFAR) algorithm is applied on each focus plane of the 3D-GPSAR image. As shown in Fig.~\ref{fig:vv_det}, the result of this procedure can be displayed as a 3D-image in which the detections are color-coded depending on the focus plane.  
At the cost of six false alarms, 11 of the 15 targets could be detected with this simple method. 
The reflections of the targets~2, 9, and 15 are visible in principle in the GPSAR image (see Fig.\ref{fig:csar_vv}), but due to clutter in close proximity they can not be detected using the implemented CA-CFAR algorithm.

In addition, the result indicates that the targets are best visible from a certain angle range (depends on the geometry). Ideally, focusing on a wrong plane would blur the reflection of point-like targets on a circle \cite{Moore2007}. But this measurement shows that the target position is shifted linearly versus the processing depth in the direction in which the target reflects the maximum power, e.g., target~7 is best seen from north-east.

Using the horizontal polarization, 9 of 15 targets (targets~1, 3, 4, 5, 7, 8, 10, 11, and 14) could be detected. At this point it should be noted that target~6 was not visible in any GPSAR image, which shows that one sensor principle alone cannot detect all different types of landmines and that several principles and a sensor fusion are required. A statistical statement regarding the detection or false alarm rate makes no sense due to the limited number of measurements and the very different types of mines used for this measurements.
\section{Conclusion}
This paper demonstrates the potential of CSAR for UAS-based detection of buried anti-personnel landmines. 11 of the 15 different test targets were successfully detected, and their position in space could be determined with an accuracy in the \si{\cm}~range. It could be shown that objects on the surface, directly below the surface as well as deeper buried objects can be detected using a side-looking geometry. Only specific tests in a realistic environment (mine type, soil composition, topography, moisture, vegetation) can generate reliable statistical statements. These effects must be systematically examined in the future in order to create a database and improve the detection algorithms.
\section{Acknowledgement}
The authors would like to thank the Urs Endress Foundation for their financial support and the German Aerospace Center in Oberpfaffenhofen making these measurements possible.
\bibliographystyle{IEEEtran}
\bibliography{schartel_arxiv} 
\end{document}